# Multiple Fizgerald-Lorentz contractions

**J.F. Geurdes**

**C. van der Lijnstraat 164**

**2593 NN Den Haag**

**Netherlands**

In a recent paper, Harada and Sachs suggested that the Fitzgerald-Lorentz contraction does not refer to a physical contraction, but can perhaps be better understood by comparing it to the conversion of currency between different coutries; e.g. the conversion between the Japanese yen in the US dollar and vice versa. In the present paper, this interpretation of the Fitzgerald-lorentz contraction will be inspected further.

## I. Introduction.

The special theory of relativity (STR) is a well-established theory which nevertheless predicts some counter-intuitive effects. One of those counter-intuitive effects is the change of size of objects moving with a certain velocity relative to each other. Mathematically, this change in size is related to the Lorentz transformation of coordinates of an observer in space-time. The difficulty with this contraction is its physical interpretation. The main question is whether an object actually changes in size when it moves relative to a relatively stationary observer. Because in STR space and time are treated on equal footing, this question also is relavant to the twin-paradox and its physical interpretation (A.D. Allan[1]).

In their paper, Harada and Sachs[2] propose to treat the Fitzgerald-Lorentz contraction similar to the change in currency between the money system of two different coutries. This point is inspected by making use of two observers. In the present paper, a number of observers are employed and the interpretation is inspected on consistency. The conclusions reached are to the opinion of the author independent of a type of interpretation.

## II. Multiple observers.

In the interpretation of Harada and Sachs, a difference is made between a unit length at relative rest and a unit length in relative motion. Suppose we inspect the situation for an arbitrary observer, $O_k$. In the view of $O_k$ we have

$$L_k = L_{kj}\sqrt{1 - \boldsymbol{b}_{kj}^2}.$$

Here, $L_k$ is the unit of length at rest in $O_k$'s frame of reference, while, $\beta_{kj}=v_{kj}/c$, is the (normed) velocity of observer $O_j$'s frame relative $O_k$'s frame and $L_{kj}$ is the unit of length in motion, relatiev to observer $O_k$ and carried by observer $O_j$. Harada and Sachs state that the factor, $(1-\beta_{kj}^2)^{1/2}$ can be compared to a 'currency conversion'. To be more specific, when $O_k$ wants to measure length in $O_j$'s units, -who moves with $\beta_{kj}$ relative to $O_k$-, he has to make use of the (currency) conversion, $(1-\beta_{kj}^2)^{1/2}$. The observer, $O_j$, measures with $L_j$, but $O_k$ has to use, $L_{kj}$, as 'currency'.

Introducing more than two observers, $O_k$ and $O_j$, we may introduce observer $O_i$ and note that, $L_k=L_{ki}(1-\beta^2_{ki})^{1/2}$. If, furthermore, $1\geq\beta_{ki}>\beta_{kj}\geq 0$, we may write for $L_{ki}$ and $L_{kj}$

$$L_{kj} = L_{ki}\sqrt{1 - b^2_{ki;j}}$$

with $\beta^2_{ki;j}$ denoting the 'velocity'

$$b^2_{ki;j} = \frac{b^2_{ki} - b^2_{kj}}{1 - b^2_{kj}}.$$

Note that, $1-\beta^2_{kj}\geq\beta^2_{ki}-\beta^2_{kj}$, hence, $\beta^2_{ki;j}$ in [0,1].

In addition to the $O_k$ 'point of view' we may also employ the $O_j$ 'point of view'. In this case we write, $L_j=L_{jk}(1-\beta^2_{jk})^{1/2}$, where $L_j$ is the unit of length at rest in $O_j$'s frame, $\beta_{jk}$ the velocity of $O_j$ relative $O_k$ and, $L_{jk}$ the 'currency' of $O_j$ when he wants to measure in $O_k$'s unit.

Because, $\beta_{jk}=\beta_{kj}$, we only may have, $L_{jk}=L_{kj}$, when, $L_j=L_k$.

Furthermore, let us state that $L_j$ is unequal to $L_k$, ($|L_j-L_k|>0$) and introduce additional observers, $O_p$, $O_q$ and $O_r$. In this case we may write, $L_p=L_{pq}(1-\beta^2_{pq})^{1/2}$ together with $L_p=L_{pr}(1-\beta^2_{pr})^{1/2}$. Hence, when the inequality $1-\beta^2_{pr}\geq\beta^2_{pq}-\beta^2_{pr}$, is valid then it follows, $\beta^2_{pq;r}$ in [0,1].

Starting from the supposition that $L_j$ is unequal to $L_k$, we are still allowed to suppose that $\beta_{pq;r}=\beta_{ki;j}$. Hence, when, $\beta_{pq}$ unequal to $\beta_{ki}$ and $\beta_{pr}$ unequal to $\beta_{kj}$, we still may have

$$\frac{b^2_{pq} - b^2_{pr}}{1 - b^2_{pr}} = \frac{b^2_{ki} - b^2_{kj}}{1 - b^2_{kj}}$$

is possible. This implies that

$$\frac{L_{pr}}{L_{pq}} = \frac{L_{kj}}{L_{ki}}$$

which ends in a contradiction when, $L_{pr}=L_{jk}$ and, $L_{pq}=L_{ki}$, because, it cannot be avoided that, given, $|L_j-L_k|>0$, $L_{kj}=L_{jk}$.

### III. Conclusion and discussion.

In this paper, we arrived at a contradiction when we started with a particular interpretation of the Fitzgerald-Lorentz contraction. The contradiction appears to be too straightforward to be dependent on a single specific interpretation. The possibility of the unearthed contradiction is demonstrated with a numercial study. E.g. given $\beta_{kj}=0.1438$, $\beta_{ki}=0.8833$, together with, $\beta_{pr}=0.0363$, $\beta_{pq}=0.8808$, we have within good approximation, $\beta_{ki;j}=\beta_{pq;r}$. Starting from $L_k=1$, we see that with the previous parameters, $L_{kj}=1.0105$, $L_{ki}=2.1329$, together with, $L_{pr}=1.0098$, $L_{pq}=2.1329$, such that within good approximation, $(L_{pr}/L_{pq})=(L_{kj}/L_{ki})$. Given this, we find that, $L_j=0.9993$.

Moreover, an alternative route is to note that $L_p = L_{jk}(1-\beta^2_{pr})^{1/2}$, leads to

$$\beta^2_{pr} = 1 - (L_p / L_{jk})^2$$

Moreover, $L_p = L_{ki}(1-\beta^2_{pq})^{1/2}$, leading to

$$\beta^2_{pq} = 1 - (L_p / L_{ki})^2$$

From the previous two expressions we find that,

$$\beta^2_{ki;j} = 1 - (L_{kj} / L_{ki})^2$$

together with

$$\beta^2_{pq;r} = 1 - (L_{jk} / L_{ki})^2$$

such that when, $\beta_{pq;r}=\beta_{ki;j}$, it easily follows that, $L_{jk}=L_{kj}$, despite $L_j$ unequal to $L_k$.

Physically, the contradiction can, for instance, occur in the field of relativistic gas dynamics[3] or, more generally, in relativistic hydrodynamics[4]. Here, the 'unit of length at rest' can be related to 'the radius of the particles at rest' that make up the gas. We may have, $L_p$, for the radius of $O_p$ particles, $L_j$, for $O_j$ particles and $L_k$, for $O_k$ particles.

If, on the other hand, the contradiction does not occur in the physical reality of, for instance, a relativistic gas, there has to be an implicit physical reason why $\beta_{pq;r}$ cannot be equal to $\beta_{ki;j}$ for unequal mutual velocities; $\beta_{kj} \neq \beta_{pr}$, $\beta_{ki} \neq \beta_{pq}$. The question then arises which kind of physical interaction we are dealing with, beacuse the contradiction is valid, even for an ideal gas.